\documentclass[%
 reprint,
]{revtex4-2}
\usepackage{graphicx}
\usepackage{dcolumn}
\usepackage{bm}
\begin{document}
\preprint{APS/123-QED}

\title{Bouncing Scenario in the $f(T)$ Modified Gravity Model with Dynamical System Analysis}
\author{S. Davood Sadatian}

\author{S. Mohamad Reza Hosseini}

\affiliation{Department of Physics, Faculty of Basic Sciences,
University
of Neyshabur,\\
P. O. Box 9319774446, Neyshabur, Iran\\
sd-sadatian@um.ac.ir , sd-sadatian@neyshabur.ac.ir ,
mrhosseiniy@yahoo.com}

\begin{abstract}
In this work, we investigate nonsingular bouncing cosmology within
the framework of quadratic modified teleparallel gravity described
by the nonlinear model $f(T)=T+\beta T^{2}$, where $\beta$
characterizes the quadratic torsion correction beyond the
teleparallel equivalent of General Relativity. Starting from the
pure tetrad formulation of $f(T)$ gravity in a spatially flat FLRW
background, we derive the modified Friedmann equations and
reformulate the cosmological evolution as a two-dimensional
autonomous dynamical system using suitable dimensionless variables
associated with the scalar field sector. The phase-space analysis
reveals the existence of saddle, unstable, and stable cosmological
configurations depending on the stability properties of the critical
points. In particular, the scalar field dominated solution behaves
as a late-time attractor under appropriate conditions on the
potential parameter. To establish the realization of a genuine
nonsingular bounce, we reconstruct the cosmological dynamics through
the regular scale factor, which remains finite and strictly positive
during the entire cosmic evolution. The corresponding Hubble
parameter satisfies the standard bouncing conditions $H(t_{b})=0,
\quad \dot{H}(t_{b})>0$, demonstrating a smooth transition from a
contracting phase to an expanding epoch without encountering a
spacetime singularity. Furthermore, the torsion scalar, effective
energy density, and geometrical background remain finite throughout
the bounce phase. We additionally analyze the evolution of the
deceleration parameter and the effective equation of state parameter
and show that the cosmological dynamics temporarily enters a
phantom-like regime near the bounce, allowing the effective
violation of the null energy condition required for nonsingular
bouncing behavior. The reconstructed bouncing solution is shown to
be compatible with the qualitative structure of the autonomous
dynamical system. In particular, the cosmological trajectory
asymptotically approaches the matter-dominated saddle configuration
while avoiding unstable stiff-matter solutions. Our results
demonstrate that quadratic $f(T)$ gravity provides a mathematically
self-consistent and physically viable framework for describing
regular bouncing cosmology through a unified combination of modified
teleparallel gravity, dynamical system analysis, and explicit
cosmological reconstruction.
\end{abstract}

\maketitle

\section{Introduction}
One of the fundamental problems of standard cosmology is the
existence of the initial singularity predicted by the classical Big
Bang scenario. In the standard Friedmann-Lemaitre-Robertson-Walker
(FLRW) framework, the cosmic evolution begins from a state of
divergent curvature, density, and temperature, where the classical
description of spacetime breaks down. This issue has motivated the
development of alternative cosmological scenarios capable of
providing a nonsingular description of the early Universe. Among
these alternatives, bouncing cosmology has attracted considerable
attention as a framework in which the Universe undergoes a smooth
transition from a contracting phase to an expanding phase without
encountering a spacetime
singularity~\cite{11a,12a,13a,14a,15a,16a,17a}.\\
Modified theories of gravity provide a natural theoretical setting
for realizing nonsingular bouncing solutions. In this context,
teleparallel gravity and its extensions have become important
candidates for describing the gravitational interaction through
torsion rather than curvature. In teleparallel gravity, gravitation
is encoded in the torsion scalar $T$, and the corresponding
generalization known as $f(T)$ gravity extends the teleparallel
equivalent of General Relativity by replacing the torsion scalar in
the gravitational action with an arbitrary function
$f(T)$~\cite{1a,2a,3a,4a,5a,6a,7a,8a,19c,20c,21c,22c,23c,24c,25c}.
An important advantage of $f(T)$ gravity is that the resulting field
equations remain second-order differential equations, in contrast to
many higher-order modified gravity theories~\cite{9a}. Important
developments in modified teleparallel gravity include the covariant
formulation of $f(T)$ gravity, the study of Lorentz invariance and
extra degrees of freedom, cosmological perturbation theory, and
nonsingular bouncing solutions~\cite{1z,2z,3z,4z,5z}.\\
The cosmological implications of $f(T)$ gravity have been widely
investigated in recent years, including late-time cosmic
acceleration, inflationary dynamics, dark energy models, compact
astrophysical objects, and nonsingular cosmological
solutions~\cite{7a,8a,9a}. In particular, nonlinear torsion
corrections can generate modifications to the Friedmann equations
that allow the violation of the effective null energy condition
required for a cosmological bounce. Consequently, teleparallel
modified gravity offers a promising framework for constructing
regular bouncing cosmologies without introducing exotic matter
components~\cite{8a}.\\
In the present work, we focus on the quadratic teleparallel gravity
model $f(T)=T+\beta T^{2},$ where the parameter $\beta$
characterizes the nonlinear torsion correction beyond the
teleparallel equivalent of General Relativity. This model provides a
simple but nontrivial extension of standard teleparallel gravity and
allows the construction of a self-consistent bouncing cosmological
scenario~\cite{ft}.\\
To investigate the global cosmological dynamics of the model, we
employ the dynamical systems approach, which provides a powerful
method for analyzing the qualitative behavior of cosmological
equations without requiring exact analytical
solutions~\cite{18a,19a,20a,21a,22a,8c,9c,10c,11c,12c}. By rewriting
the cosmological field equations as an autonomous system, one can
identify the critical points of the phase space and study their
stability properties. This method is particularly useful for
understanding the asymptotic behavior of bouncing cosmologies and
their corresponding cosmological phases.\\
The novelty of the present analysis lies in the development of a
single self-consistent framework of quadratic teleparallel gravity
based on the nonlinear model $f(T)=T+\beta T^{2}$. All cosmological
quantities, including the effective torsion density and pressure,
the autonomous dynamical system, and the bouncing cosmological
solution, are derived from the modified $f(T)$ field equations
without introducing phenomenological assumptions. The cosmological
bounce is reconstructed explicitly through a regular scale factor,
allowing the verification of the bounce conditions $H(t_{b})=0$,
$\dot{H}(t_{b})>0$, while ensuring the regularity of the torsion
scalar and the positivity of the scale factor throughout the cosmic
evolution. The reconstructed bouncing solution is shown to be
compatible with the qualitative structure of the autonomous
dynamical system. In particular, the cosmological trajectory
asymptotically approaches the matter-dominated saddle configuration
while avoiding unstable stiff-matter solutions. Furthermore,
analytic expressions for the scale factor, Hubble parameter,
deceleration parameter, and effective equation of state parameter
demonstrate that the cosmological dynamics remain regular and
physically consistent
within a modified gravity scenario.\\
The structure of the paper is organized as follows. In Section 2, we
present the formulation of the quadratic $f(T)$ gravity model and
derive the corresponding modified Friedmann equations. In Section 3,
we construct the autonomous dynamical system and investigate the
stability properties of the critical points. In Sections 4 we
reconstruct the nonsingular bouncing solution and analyze the
evolution of the cosmological parameters. Finally, in Section 5, we
summarize the
main results.\\

\section{Gravity Model}
In this model, we consider the gravitational action as a function of
the torsion scalar $T$ (in the teleparallel framework), which is
minimally coupled to a scalar field $\varphi$ with a kinetic term
and potential~\cite{ft,7a}. In the teleparallel formulation of
gravity, the dynamical variables are the tetrad fields $e^{A}_{\
\mu}$, which satisfy
\begin{equation}
g_{\mu\nu}=\eta_{AB}e^{A}_{\ \mu}e^{B}_{\ \nu},
\end{equation}
where $\eta_{AB}=\mathrm{diag}(1,-1,-1,-1)$ is the Minkowski metric
in the tangent space. The determinant of the tetrad is denoted by
\begin{equation}
e=\det(e^{A}_{\ \mu})=\sqrt{-g}.
\end{equation}
In teleparallel gravity, the gravitational interaction is described
by torsion rather than curvature through the Weitzenb\"ock
connection
\begin{equation}
\Gamma^{\lambda}_{\ \mu\nu} = e_{A}^{\
\lambda}\partial_{\nu}e^{A}_{\ \mu}.
\end{equation}
The torsion tensor is defined as
\begin{equation}
T^{\lambda}_{\ \mu\nu} = \Gamma^{\lambda}_{\ \nu\mu} -
\Gamma^{\lambda}_{\ \mu\nu},
\end{equation}
while the contorsion tensor is
\begin{equation}
K^{\mu\nu}_{\ \ \rho} = -\frac{1}{2} \left( T^{\mu\nu}_{\ \ \rho} -
T^{\nu\mu}_{\ \ \rho} - T_{\rho}^{\ \mu\nu} \right).
\end{equation}
The superpotential tensor is given by
\begin{equation}
S_{\rho}^{\ \mu\nu} = \frac{1}{2} \left( K^{\mu\nu}_{\ \ \rho} +
\delta^{\mu}_{\rho}T^{\alpha\nu}_{\ \ \alpha} -
\delta^{\nu}_{\rho}T^{\alpha\mu}_{\ \ \alpha} \right).
\end{equation}
The torsion scalar is therefore defined as
\begin{equation}
T = S_{\rho}^{\ \mu\nu} T^{\rho}_{\ \mu\nu}.
\end{equation}
Therefore, the action of the $f(T)$ gravity minimally coupled to a
scalar field is written as
\begin{equation}
S = \int d^{4}x \ e \left[ \frac{1}{2\kappa^{2}}f(T) +
\frac{1}{2}\partial_{\mu}\varphi \partial^{\mu}\varphi - V(\varphi)
\right], \label{actionnew}
\end{equation}
where $\kappa^{2}=8\pi G$ is the gravitational coupling constant. In
the present work, we adopt the pure tetrad formulation of $f(T)$
gravity. For the spatially flat FLRW background,
\begin{equation}
ds^{2}=dt^{2}-a(t)^{2}\delta_{ij}dx^{i}dx^{j},
\end{equation}
we choose the diagonal tetrad
\begin{equation}
e^{A}_{\ \mu} = \mathrm{diag}(1,a(t),a(t),a(t)).
\end{equation}
For this tetrad, the inertial spin connection vanishes consistently,
and the torsion scalar becomes
\begin{equation}\label{torsionH}
T=-6H^{2},
\end{equation}
where $H=\dot a/a$ is the Hubble parameter (Throughout the present
work, we adopt the metric signature $(+,-,-,-)$, which consistently
leads to the Eq.~(\ref{torsionH}) teleparallel relation for the
diagonal flat FLRW tetrad, and all modified Friedmann equations and
cosmological quantities are derived consistently using the
convention given in Eq.~(\ref{torsionH})). By varying the action
(\ref{actionnew}) with respect to the tetrad field, the modified
Friedmann equations are obtained as
\begin{equation}
12H^{2}f_{T}+f(T) = 2\kappa^{2}\rho_{\varphi},
\end{equation}
\begin{equation}
48H^{2}\dot H f_{TT} - 4(\dot H+3H^{2})f_{T} - f(T) =
2\kappa^{2}P_{\varphi},
\end{equation}
where $f_{T}=df/dT$ and $f_{TT}=d^{2}f/dT^{2}$. The scalar field
energy density and pressure are
\begin{equation}
\rho_{\varphi} = \frac{1}{2}\dot{\varphi}^{2}+V(\varphi),
\end{equation}
\begin{equation}
P_{\varphi} = \frac{1}{2}\dot{\varphi}^{2}-V(\varphi).
\end{equation}
In modified teleparallel gravity, the gravitational action is
generalized from the Teleparallel Equivalent of General Relativity
(TEGR) by replacing the torsion scalar $T$ with an arbitrary
function $f(T)$. Nonlinear extensions of the form $f(T)\neq T$ can
lead to modified cosmological dynamics while preserving the
second-order nature of the field equations, in contrast to many
higher-order modified gravity theories. In the present work, we
focus on the quadratic model $f(T)=T+\beta T^{2}$, where
the parameter $\beta$ characterizes the nonlinear torsion correction beyond TEGR.
This simple but nontrivial extension provides a consistent framework for
investigating nonsingular bouncing cosmology within modified teleparallel gravity.\\
To establish a direct connection between the cosmological dynamics
and the underlying modified teleparallel gravity theory, we consider
the explicit quadratic form~\cite{ft}
\begin{equation}
f(T)=T+\beta T^{2}, \label{ftmodel}
\end{equation}
where $\beta$ is a constant parameter controlling the nonlinear
torsion correction. Using the modified Friedmann equations, the
torsion contribution can be interpreted as an effective fluid sector
with density and pressure given by
\begin{equation}
\rho_{T} = \frac{1}{2\kappa^{2}} \left( 2Tf_{T}-f-T \right),
\label{rhoT}
\end{equation}
and
\begin{equation}
P_{T} = -\frac{1}{2\kappa^{2}} \left[ 4\dot{H} \left(
2Tf_{TT}+f_{T}-1 \right) + \rho_{T} \right]. \label{PT}
\end{equation}
For the model (\ref{ftmodel}), one obtains
\begin{equation}
f_{T}=1+2\beta T,
\end{equation}
\begin{equation}
f_{TT}=2\beta.
\end{equation}
Using the relation between torsion and the Hubble parameter Eq.
(\ref{torsionH}), the effective torsion density becomes
\begin{equation}
\rho_{T} = \frac{54\beta}{\kappa^{2}}H^{4},
\end{equation}
while the effective torsion pressure is
\begin{equation}
P_{T} = -\frac{18\beta}{\kappa^{2}} H^{2} \left( 4\dot{H}+3H^{2}
\right).
\end{equation}
In the following, the effective torsion sector used in the dynamical
analysis is not introduced phenomenologically, but follows directly
from the chosen nonlinear $f(T)$ gravity model~\cite{ft}.\\

\section{Dynamical System Analysis}
The dynamical systems approach provides a powerful framework for
investigating the qualitative behavior of cosmological models
without requiring exact analytical solutions. By rewriting the
cosmological field equations as an autonomous system, one can
identify the critical points of the phase space and analyze their
stability properties. This method is particularly useful for
determining the asymptotic behavior of cosmological solutions and
studying the transition between different evolutionary phases, such
as contracting, expanding, or bouncing regimes. In the present work,
the dynamical analysis is employed to investigate the stability
structure of the quadratic $f(T)=T+\beta T^{2}$ cosmological model
and to identify the corresponding physically relevant cosmological
solutions. It should be emphasized that the autonomous system considered
in the present work describes the effective cosmological dynamics
after incorporating the quadratic torsion corrections into the
modified Friedmann constraint. Consequently, the nonlinear $f(T)$
contributions affect the phase-space structure indirectly through
the modified background evolution.\\
In order to investigate the cosmological dynamics of the model, we
introduce the following dimensionless variables~\cite{df}
\begin{equation}
x=\frac{\dot{\varphi}}{\sqrt{6}H}, \qquad
y=\frac{\sqrt{V(\varphi)}}{\sqrt{3}H}. \label{dynvars}
\end{equation}
These variables are widely used in scalar field cosmology and allow
the cosmological equations to be rewritten as an autonomous
dynamical system. Using the modified Friedmann equation, the density
parameter associated with the scalar field becomes
\begin{equation}
\Omega_{\varphi}=x^{2}+y^{2}.
\end{equation}
For the exponential potential
\begin{equation}
V(\varphi)=V_{0}e^{-\lambda \varphi}, \label{expotential}
\end{equation}
where the slope parameter of the scalar field potential
\begin{equation}
\lambda=-\frac{1}{V}\frac{dV}{d\varphi}
\end{equation}
is constant and dimensionless. The exponential potential
$V(\varphi)=V_{0}e^{-\lambda \varphi}$, is widely used in
cosmological dynamical system analysis because it leads to a closed
autonomous system with a constant parameter
$\lambda=-\frac{1}{V}\frac{dV}{d\varphi}$. This property
considerably simplifies the phase-space structure and allows a
systematic investigation of the critical points and their stability
properties.\\
The scalar field satisfies the Klein-Gordon equation
\begin{equation}
\ddot{\varphi}+3H\dot{\varphi}+\frac{dV}{d\varphi}=0. \label{KG}
\end{equation}
Using the number of e-foldings
\begin{equation}
N=\ln a,
\end{equation}
together with Eqs. (\ref{dynvars}),(\ref{KG}), the cosmological
equations can be written as the autonomous system (using
$\frac{d}{dN}=\frac{1}{H}\frac{d}{dt}$, together with the scalar
field equation and modified Friedmann equations)
\begin{equation}
\frac{dx}{dN} = -3x +\sqrt{\frac{3}{2}}\lambda y^{2} +\frac{3}{2}x
\left( 2x^{2} +\gamma(1-x^{2}-y^{2}) \right), \label{auto1}
\end{equation}

\begin{equation}
\frac{dy}{dN} = -\sqrt{\frac{3}{2}}\lambda xy +\frac{3}{2}y \left(
2x^{2} +\gamma(1-x^{2}-y^{2}) \right), \label{auto2}
\end{equation}
where $\gamma$ denotes the barotropic index of the background matter
fluid.  The physical phase space is constrained by
\begin{equation}
0\le x^{2}+y^{2}\le 1.
\end{equation}
The effective equation of state parameter of the scalar field is
given by
\begin{equation}
\omega_{\varphi} = \frac{x^{2}-y^{2}}{x^{2}+y^{2}},
\end{equation}
while the total effective equation of state becomes
\begin{equation}
\omega_{\rm eff} = x^{2}-y^{2} +\gamma(1-x^{2}-y^{2}).
\end{equation}
The evolution of the effective equation of state parameter is
illustrated in Fig.~1, which shows the transition between different
cosmological regimes in the quadratic $f(T)$ gravity model (For the
numerical reconstruction of the effective equation of state
parameter shown in Fig.~1, we adopted the representative parameter
values $\gamma = 1$, corresponding to a pressureless matter
background. The phase-space variables were defined in the interval
$-1 \le x \le 1, \quad -1 \le y \le 1$, subject to the physical
constraint $ 0 \le x^{2}+y^{2}\le 1$).\\
Using the dimensionless variables introduced above, the cosmological
equations can be reformulated as a two-dimensional autonomous system
whose equilibrium points characterize the asymptotic cosmological
behavior of the model. The corresponding stability analysis is
performed through the eigenvalues of the Jacobian matrix associated
with the autonomous system.\\

\subsection{Critical Points and Stability Analysis}
The critical points of the autonomous system are obtained from the
conditions
\begin{equation}
\frac{dx}{dN}=0, \qquad \frac{dy}{dN}=0.
\end{equation}
The corresponding Jacobian matrix is
\begin{equation}
J = \left(
\begin{array}{cc}
\frac{\partial f}{\partial x} & \frac{\partial f}{\partial y} \\[2pt]
\frac{\partial g}{\partial x} & \frac{\partial g}{\partial y}
\end{array}
\right),
\end{equation}
where $f(x,y)$ and $g(x,y)$ denote the right-hand sides of Eqs.
(\ref{auto1}) and (\ref{auto2}). The eigenvalues of the Jacobian
matrix determine the local stability properties of the critical
points. Stable nodes correspond to eigenvalues with negative real
parts, unstable nodes correspond to positive eigenvalues, while
saddle points possess eigenvalues with opposite signs. The
physically relevant critical points of the system are summarized in
Table~I. It is important to note that although the autonomous system
does not explicitly contain $\beta$, the quadratic torsion parameter
enters through the modified Friedmann constraint:
$x^2+y^2=1-\frac{54\beta}{\kappa^2} H^2$. Thus, $\beta$ implicitly
restricts the accessible phase space and affects the global
dynamics, especially near the bounce where $H\to 0$. This indirect
role is sufficient for the stability analysis of critical points, as
the latter are defined at the level of the autonomous equations,
while the $\beta$-dependence reappears in the reconstruction of
physical trajectories.\\
\begin{table}[h]
\centering \caption{Critical points and their stability properties
for the autonomous system.}
\begin{tabular}{c c c c}
\hline
Point & $(x,y)$ & Existence & Stability \\
\hline
$P_{1}$ & $(0,0)$ & Always & Saddle point \\
$P_{2}$ & $(1,0)$ & Always & Unstable node \\
$P_{3}$ & $(-1,0)$ & Always & Unstable node \\
$P_{4}$ & $\left( \frac{\lambda}{\sqrt{6}},
\sqrt{1-\frac{\lambda^{2}}{6}} \right)$ & $\lambda^{2}<6$ & Stable
for $\lambda^{2}<3\gamma$
\\
\hline
\end{tabular}
\label{table1new}
\end{table}
To verify the stability properties of the critical points, we
explicitly evaluated the eigenvalues of the Jacobian matrix
associated with the autonomous system. The critical point $P_{1}$
corresponds to a matter-dominated phase and behaves as a saddle
point, as determined by the existence of eigenvalues with opposite
signs. The points $P_{2}$ and $P_{3}$ represent kinetically
dominated scalar field solutions with a stiff equation of state; for
both points, at least one eigenvalue becomes positive, implying that
they are unstable nodes. This instability follows directly from the
linearized system and does not require any additional
center-manifold analysis. The point $P_{4}$ corresponds to a scalar
field dominated accelerated solution. For $\lambda^{2} < 3\gamma$,
all eigenvalues of $P_{4}$ have negative real parts, demonstrating
that it behaves as a stable late-time attractor in the physical
phase space.\\
For completeness, we now present the explicit eigenvalue structure
associated with the critical points of the autonomous system. The
Jacobian matrix corresponding to the system (Eqs.~(29)-(30)) is
evaluated at each equilibrium point listed in Table~I. For the
matter-dominated critical point $P_{1}=(0,0)$, the eigenvalues are
obtained as $\lambda_{1}=\frac{3}{2}(\gamma-2), \quad
\lambda_{2}=\frac{3}{2}\gamma$. For the physically relevant range
$(0<\gamma<2)$, the eigenvalues possess opposite signs,
demonstrating that $(P_{1})$ is a saddle point. For the
kinetic-dominated solutions $P_{2}=(1,0), \quad P_{3}=(-1,0)$, the
eigenvalues are respectively $P_{2}:\quad \lambda_{1}=3(2-\gamma),
\quad \lambda_{2}=3-\sqrt{\frac{3}{2}}\lambda$, and $P_{3}:\quad
\lambda_{1}=3(2-\gamma), \quad
\lambda_{2}=3+\sqrt{\frac{3}{2}}\lambda$. At least one eigenvalue is
positive for the physically relevant parameter region, implying that
both points behave as unstable nodes. For the scalar field dominated
critical point $P_{4}= \left( \frac{\lambda}{\sqrt{6}},
\sqrt{1-\frac{\lambda^{2}}{6}} \right)$, the eigenvalues become
$\lambda_{\pm} = \frac{1}{2} \left[ (\lambda^{2}-6) \pm \sqrt{
(\lambda^{2}-6)^{2} + 24\lambda^{2} - 24\gamma\lambda^{2} } \right].
$ The condition $\lambda^{2}<3\gamma$ guarantees that the real parts
of both eigenvalues remain negative, confirming that $(P_{4})$
corresponds to a stable late-time attractor. Therefore, the
stability classification summarized in Table~I follows directly from
the explicit eigenvalue structure of the Jacobian matrix and does
not require additional center-manifold analysis.\\
The critical points of the autonomous system listed in Table~I are
explicitly indicated in Fig.~1. The point $P_{1}=(0,0)$ corresponds
to a matter-dominated saddle configuration. The kinetic-dominated
scalar field solutions $P_{2}=(1,0)$ and $P_{3}=(-1,0)$ behave as
unstable nodes with stiff-fluid behavior. The point $P_{4} = \left(
\frac{\lambda}{\sqrt{6}}, \sqrt{1-\frac{\lambda^{2}}{6}} \right)$,
represents the scalar field dominated accelerated solution, which
becomes a stable late-time attractor for $\lambda^{2}<3\gamma$. The
shaded region with $\omega_{\rm eff}<-1/3$ corresponds to
accelerated cosmological expansion, while the limit $\omega_{\rm
eff}\rightarrow -1$ describes the asymptotic approach toward a
dark-energy dominated phase.\\
Since the exponential potential leads to a constant parameter
$\lambda$, the cosmological dynamics is completely described by the
two-dimensional autonomous system (\ref{auto1})-(\ref{auto2}).
Therefore, the physical phase space is fully represented in the
$(x,y)$ plane, and no additional dynamical variable is required.
Consequently, the phase portraits shown in this work correspond to
the complete dynamical evolution of the system rather than to
lower-dimensional projections of a higher-dimensional phase space.
The corresponding phase-space trajectories of the autonomous system
are displayed in Fig.~2, where the global structure of the
cosmological phase space and
the stability behavior of the critical points are illustrated.\\
\begin{figure}
\centering
\includegraphics[width=0.45\textwidth, keepaspectratio]{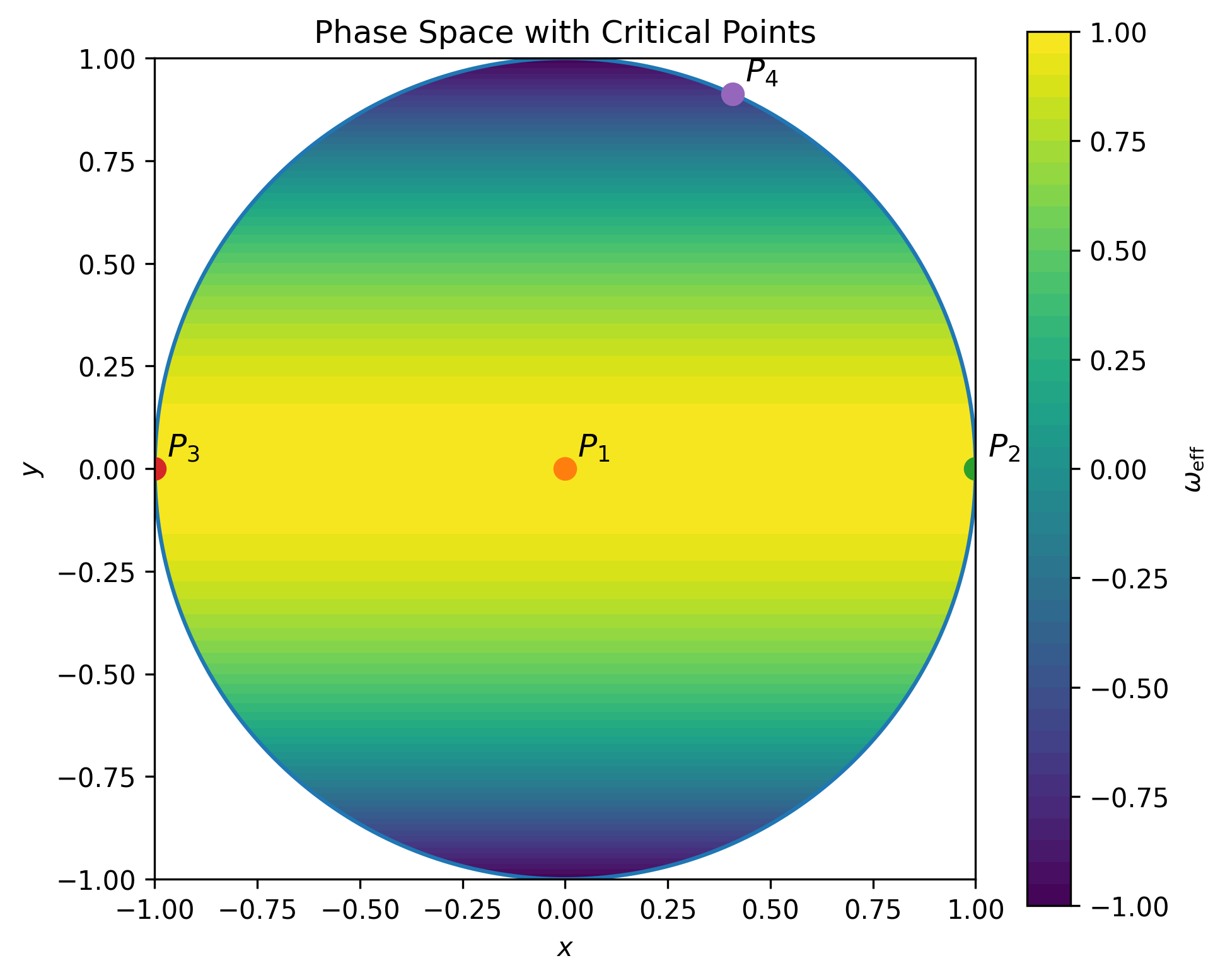}
\caption{\small Contour evolution of the effective equation of state
parameter $\omega_{\rm eff}$ in the physical phase space of the
two-dimensional autonomous system corresponding to the quadratic
$f(T)=T+\beta T^{2}$ cosmological model. The shaded region is
determined from $\omega_{\rm eff} = x^{2}-y^{2}
+\gamma(1-x^{2}-y^{2})$, using the dimensionless variables
$x=\dot{\varphi}/(\sqrt{6}H)$ and $y=\sqrt{V}/(\sqrt{3}H)$ with the
exponential potential $V(\varphi)=V_{0}e^{-\lambda\varphi}$. The
physical phase space is restricted by the condition $0\le
x^{2}+y^{2}\le1$. Regions with $\omega_{\rm eff}<-1/3$ correspond to
accelerated cosmological expansion, while $\omega_{\rm
eff}\rightarrow -1$ represents the approach toward a dark-energy
dominated phase.} \label{fig:omega_eff}
\end{figure}

\begin{figure}
\centering
\includegraphics[width=0.45\textwidth, keepaspectratio]{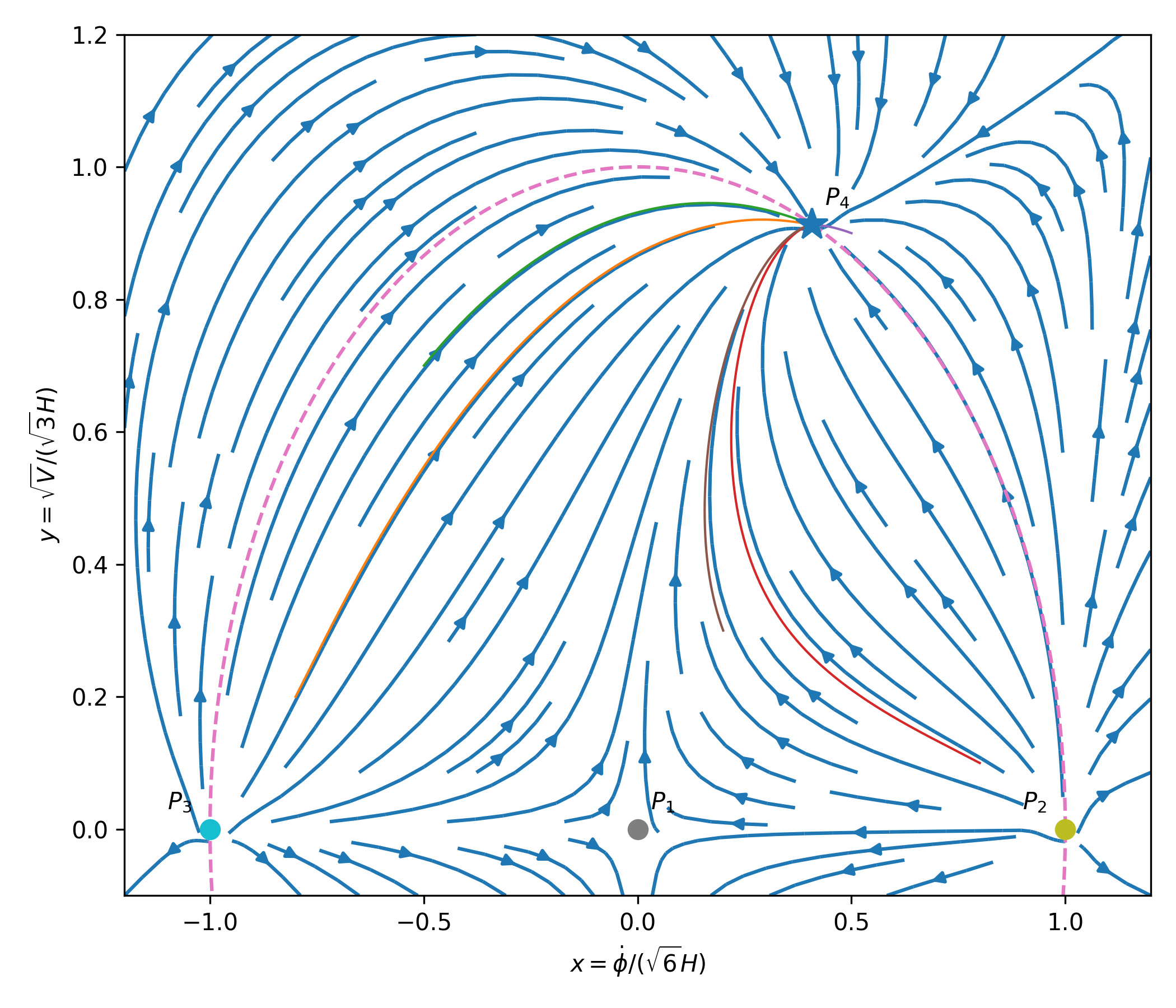}
\caption{\small Complete phase-space trajectories of the
two-dimensional autonomous system in the $(x,y)$ plane for
representative values of the model parameters. The numerical
integration is performed for the representative parameter values
$\beta=1$, $\lambda=1$, and $n=1$. The figure illustrates the global
structure of the cosmological phase space and the stability
behavior.}
\end{figure}
It is important to emphasize that the apparent divergences appearing
in some phase-space trajectories (see Fig.~2) do not correspond to
physical spacetime singularities. These features originate from the
normalization of the dynamical variables used in the autonomous
formulation, where inverse powers of the Hubble parameter may
appear. Consequently, near the bounce point where $H \rightarrow 0$,
some normalized variables can become formally divergent even though
the underlying cosmological background remains completely regular.
To verify the physical consistency of the model, we reconstruct the
cosmological evolution directly in terms of the physical quantities
$a(t)$, $H(t)$, and the torsion scalar $T = -6H^{2}$. As
demonstrated in Section~IV, the resulting evolution confirms that
the scale factor, Hubble parameter, and torsion scalar remain finite
throughout the bounce phase. This explicitly demonstrates that the
apparent phase-space divergences are artifacts of the chosen
dynamical parametrization rather than genuine physical
singularities.\\

\section{The Bouncing scenario in The $f(T)$ Model}
We employ a single consistent modified teleparallel gravity model
throughout the manuscript, namely the quadratic form $f(T)=T+\beta
T^{2}$, where $\beta$ characterizes the nonlinear torsion correction
beyond the teleparallel equivalent of General Relativity. All
cosmological quantities, including the effective torsion density and
pressure, the autonomous dynamical system, and the reconstructed
bouncing solution, are derived consistently from this underlying
action. Within this framework, the modified Friedmann equations can
admit regular nonsingular cosmological solutions in which the Hubble
parameter evolves smoothly from a contracting phase ($H<0$) to an
expanding phase ($H>0$), driven by the nonlinear torsion corrections
together with the scalar field sector. To investigate the global
behavior of the model, we reformulate the cosmological equations as
a two-dimensional autonomous dynamical system. The resulting phase
portraits allow us to identify the critical points, determine their
stability properties, and examine the dynamical evolution of
cosmological trajectories. A nonsingular bounce corresponds
precisely to a smooth transition from contraction to expansion
without encountering a spacetime singularity.\\
In this regard, the dynamical system analysis presented in the
previous section provides the global phase-space structure of the
quadratic $f(T)=T+\beta T^{2}$ cosmological model and determines the
asymptotic behavior of the corresponding cosmological solutions.
Although the bouncing solution considered in the present section is
reconstructed through an explicit ansatz for the scale factor, it
remains compatible with the qualitative dynamics identified in the
autonomous analysis. In particular, the modified Friedmann
constraint $x^{2}+y^{2} = 1-\frac{54\beta}{\kappa^{2}}H^{2}$,
establishes a direct connection between the physical phase space and
the Hubble parameter governing the bounce dynamics. Near the bounce
point, where $H\rightarrow0$, the cosmological trajectory approaches
the boundary of the physical phase space, demonstrating that the
reconstructed bounce evolution is naturally embedded within the
dynamical structure of the system. Furthermore, the late-time
accelerated behavior obtained from the reconstructed cosmological
solution is consistent with the asymptotic attractor structure
associated with the scalar field dominated critical point $P_{4}$.
Consequently, the autonomous dynamical analysis and the explicit
bounce reconstruction should be interpreted as complementary
components of a single self-consistent modified teleparallel
cosmological framework. We discuss this issue in more
detail in the subsection at the end of this section.\\
Based on the above discussion, we investigate the cosmological
implications of the quadratic teleparallel gravity
model~(\ref{ftmodel}) to examine a bouncing scenario. To reconstruct
the cosmological evolution explicitly, we consider the bouncing
scale factor
\begin{equation}
a(t)=a_{0}(1+\sigma t^{2})^{n}, \label{scalea}
\end{equation}
where $a_{0}>0$, $\sigma>0$ (dimensions of inverse time squared),
and $n>0$ are constant parameters. $a_{0}$ represents the present
normalization of the scale factor and is dimensionless. This scale
factor remains finite and strictly positive for all cosmic times and
therefore describes a nonsingular cosmological background. The
behavior of the reconstructed bouncing scale factor is presented in
Fig.~3, where the minimum value of $a(t)$ at the bounce point
demonstrates the absence of a cosmological singularity. Using Eq.
(\ref{scalea}), the Hubble parameter becomes
\begin{equation}
H(t) = \frac{\dot a}{a} = \frac{2n\sigma t}{1+\sigma t^{2}}.
\label{HubbleNew}
\end{equation}
The evolution of the Hubble parameter shows that
\begin{equation}
H<0 \quad \text{for} \quad t<0,
\end{equation}
corresponding to the contracting phase, while
\begin{equation}
H>0 \quad \text{for} \quad t>0,
\end{equation}
describes the expanding epoch after the bounce. At the bounce point
$t=0$, we obtain
\begin{equation}
H(0)=0,
\end{equation}
which confirms the smooth transition between contraction and
expansion. The corresponding evolution of the Hubble parameter is
shown in Fig.~4, which explicitly demonstrates the transition from
the contracting phase to the expanding epoch through the bounce
point. The time derivative of the Hubble parameter is
\begin{equation}
\dot H(t) = \frac{2n\sigma(1-\sigma t^{2})} {(1+\sigma t^{2})^{2}}.
\label{HdotNew}
\end{equation}
At the bounce point,
\begin{equation}
\dot H(0)=2n\sigma>0,
\end{equation}
which guarantees the occurrence of a genuine nonsingular bounce. The
torsion scalar associated with the teleparallel geometry becomes
\begin{equation}
T=-6H^{2} = -\frac{24n^{2}\sigma^{2}t^{2}} {(1+\sigma t^{2})^{2}},
\end{equation}
which remains finite throughout the cosmic evolution, which confirms
the regularity of the teleparallel background geometry at the bounce
point. Using the quadratic form of the modified gravity function,
$f(T)=T+\beta T^{2}$, the nonlinear torsion correction becomes
\begin{equation}
f(T) = -\frac{24n^{2}\sigma^{2}t^{2}} {(1+\sigma t^{2})^{2}} + \beta
\left( \frac{24n^{2}\sigma^{2}t^{2}} {(1+\sigma t^{2})^{2}}
\right)^{2}.
\end{equation}
This expression demonstrates explicitly how the modified
gravitational sector evolves during the cosmological bounce and
shows that the nonlinear torsion contribution remains finite
throughout the entire cosmic evolution. Consequently, Eq.~(44)
confirms the regularity of the modified teleparallel background and
verifies that the bounce is generated dynamically by the quadratic
torsion correction rather than by introducing phenomenological
modifications at the level of the scale factor alone. Moreover,
Eq.~(44) provides the explicit form of the modified gravity function
entering the Friedmann equations during the bounce phase. This
allows one to analyze how the quadratic torsion term contributes
near the bounce point, where the deviations from the teleparallel
equivalent of General Relativity become significant. In particular,
the $\beta T^{2}$ correction dominates the high-energy regime around
$H \rightarrow 0$, thereby playing a central role in realizing the
effective violation of the null energy condition required for a
nonsingular cosmological bounce. Therefore, Eq.~(44) is not merely a
formal substitution, but an explicit reconstruction of the modified
gravitational dynamics associated with the bouncing solution. It
demonstrates the internal consistency between the reconstructed
cosmological background and the underlying quadratic $f(T)$ gravity
framework adopted throughout the manuscript.\\
The deceleration parameter is defined as~\cite{de}
\begin{equation}
q = -1-\frac{\dot H}{H^{2}}.
\end{equation}
Using Eqs. (\ref{HubbleNew}) and (\ref{HdotNew}), we obtain
\begin{equation}
q(t) = -1 - \frac{ 1-\sigma t^{2} } { 2n\sigma t^{2} }. \label{qnew}
\end{equation}
The behavior of the deceleration parameter indicates that the
Universe undergoes a transition from decelerated contraction to
accelerated expansion around the bounce point. The effective
equation of state parameter is given by
\begin{equation}
\omega_{\rm eff} = -1 - \frac{2\dot H}{3H^{2}}.
\end{equation}
Substituting Eqs. (\ref{HubbleNew}) and (\ref{HdotNew}) yields
\begin{equation}
\omega_{\rm eff}(t) = -1 - \frac{ 1-\sigma t^{2} } { 3n\sigma t^{2}
}. \label{omeganew}
\end{equation}
According to the above relations, the null energy condition is
effectively violated near the bounce region since $\rho_{\rm
eff}+P_{\rm eff}<0$, which follows directly from the relation $\dot
H=-\frac{\kappa^2}{2}(\rho_{\rm eff}+P_{\rm eff})$, together with
the bounce condition $\dot H>0$ at the
bouncing time $t=t_b$.\\
The evolution of the effective equation of state parameter
associated with the reconstructed bounce solution is displayed in
Fig.~5. It is important to emphasize that the apparent divergence of
the deceleration parameter and the effective equation of state
parameter at the bounce point does not correspond to a physical
spacetime singularity. Both quantities are defined through inverse
powers of the Hubble parameter, $q \propto H^{-2}$ and $\omega_{\rm
eff} \propto H^{-2}$, and therefore become formally ill-defined at
the bounce where $H=0$. However, the fundamental geometrical and
physical quantities, including the scale factor, Hubble parameter,
torsion scalar, and effective energy density, remain finite and
regular throughout the cosmic evolution. Consequently, the
divergence of $q$ and $\omega_{\rm eff}$ should be interpreted as a
limitation of the effective-fluid parametrization rather than a
genuine cosmological singularity. The evolution of the deceleration
parameter corresponding to the reconstructed bouncing solution is
presented in Fig.~\ref{fig:qparameter}. The figure demonstrates the
transition from a contracting decelerated epoch toward an
accelerated expanding phase after the bounce. Also, the quadratic
$f(T)$ model successfully provides a self-consistent realization of
a regular bouncing Universe with all geometrical quantities such as
the scale factor, Hubble parameter, and torsion scalar remain finite
and smooth transition between contraction
and expansion phases.\\
\begin{figure}[htbp]
\centering
\includegraphics[width=0.75\linewidth]{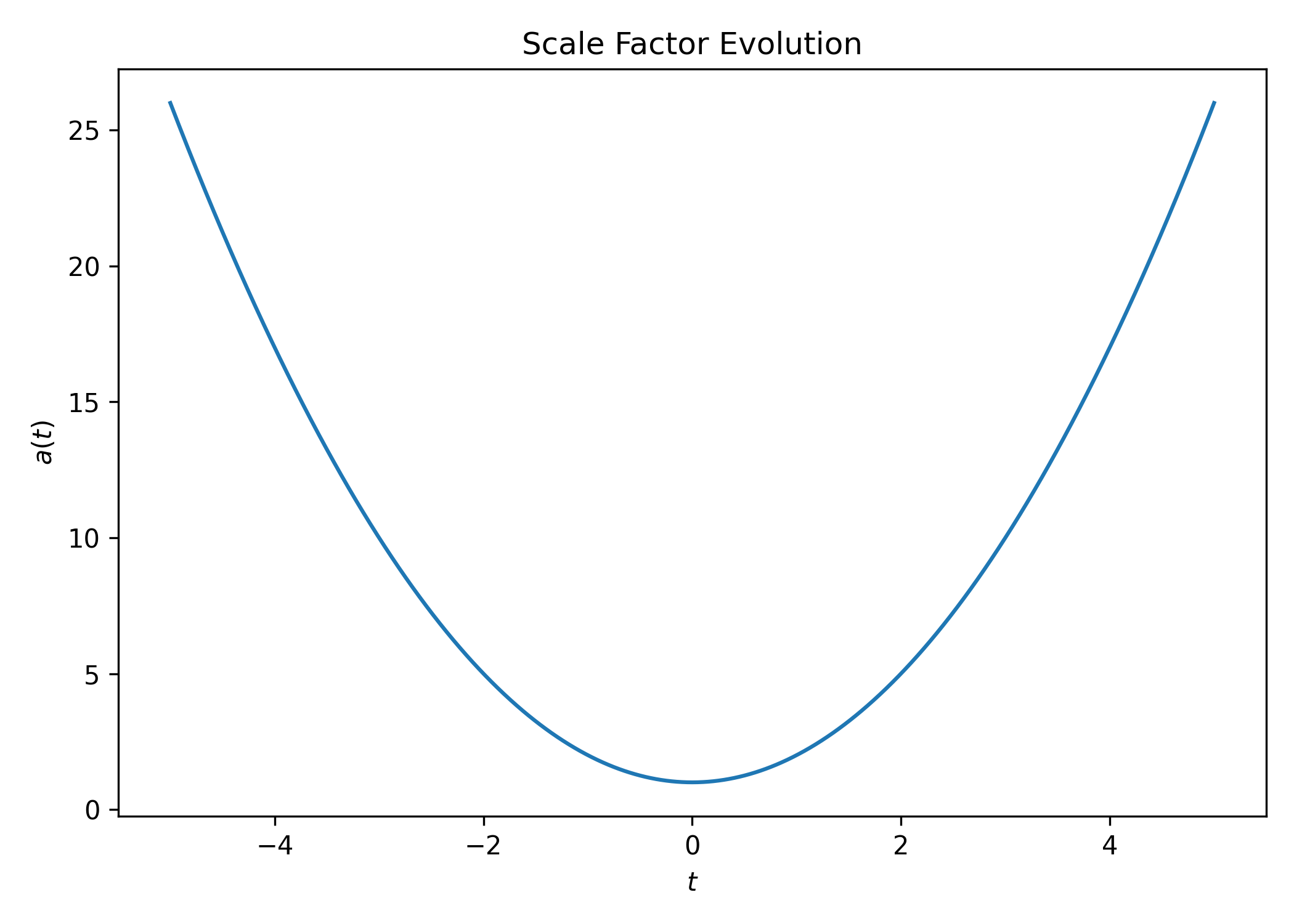}
\caption{ Evolution of the bouncing scale factor
$a(t)=a_{0}(1+\sigma t^{2})^{n}$. The scale factor remains finite
and strictly positive throughout the cosmic evolution, demonstrating
the absence of a Big Bang singularity. The minimum value of $a(t)$
occurs at the bounce point $t=0$, where the transition from
contraction to expansion takes place smoothly. }
\label{fig:scale_factor}
\end{figure}

\begin{figure}[htbp]
\centering
\includegraphics[width=0.75\linewidth]{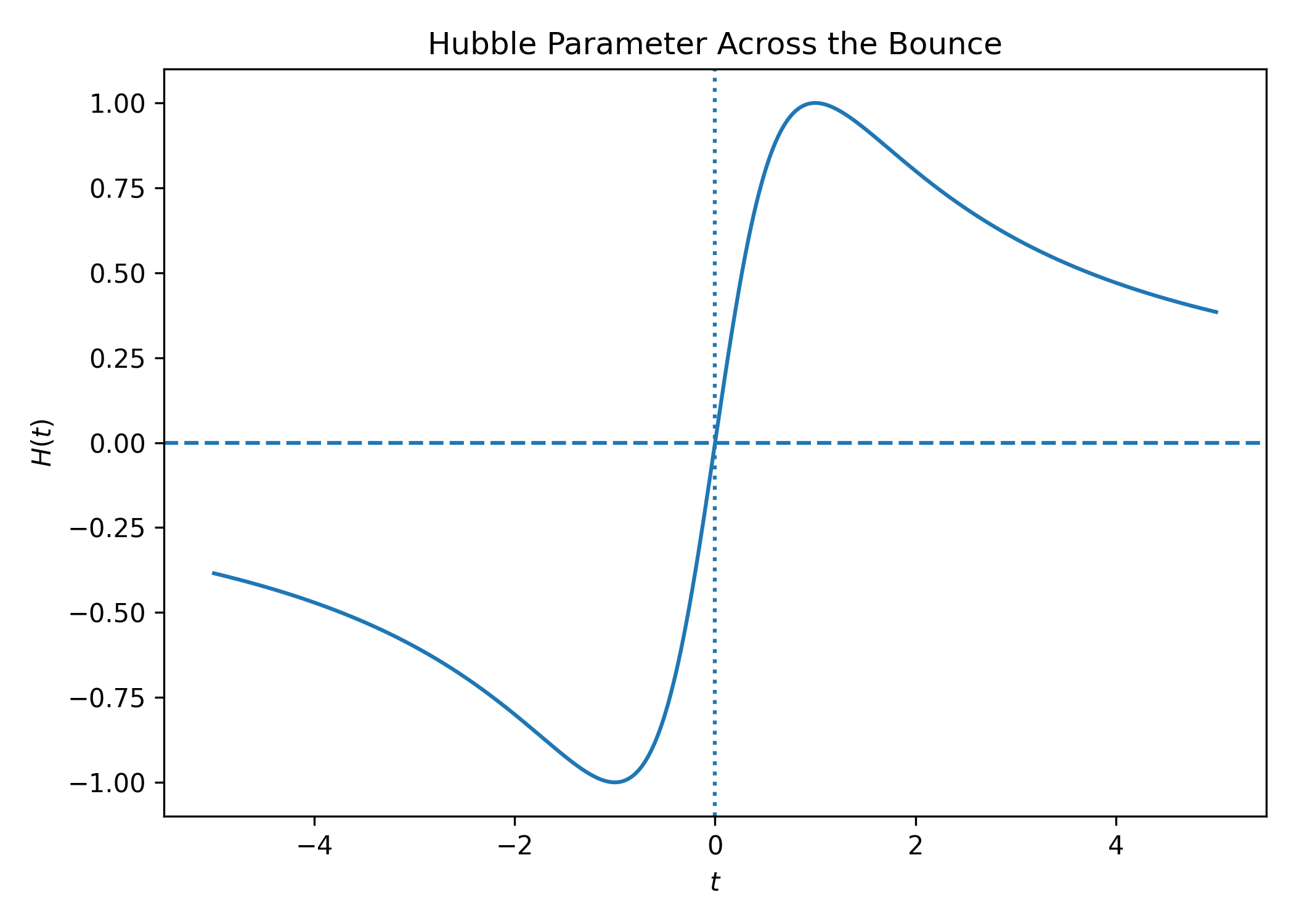}
\caption{ Evolution of the Hubble parameter $H(t)=\frac{2n\sigma
t}{1+\sigma t^{2}}$. The figure explicitly shows the transition from
the contracting phase $(H<0)$ to the expanding phase $(H>0)$ through
the bounce point at $t=0$, where $H=0$ and $\dot{H}>0$. This
behavior confirms the realization of a nonsingular cosmological
bounce within the quadratic $f(T)$ gravity framework. }
\label{fig:hubble_bounce}
\end{figure}

\begin{figure}[htbp]
\centering
\includegraphics[width=0.75\linewidth]{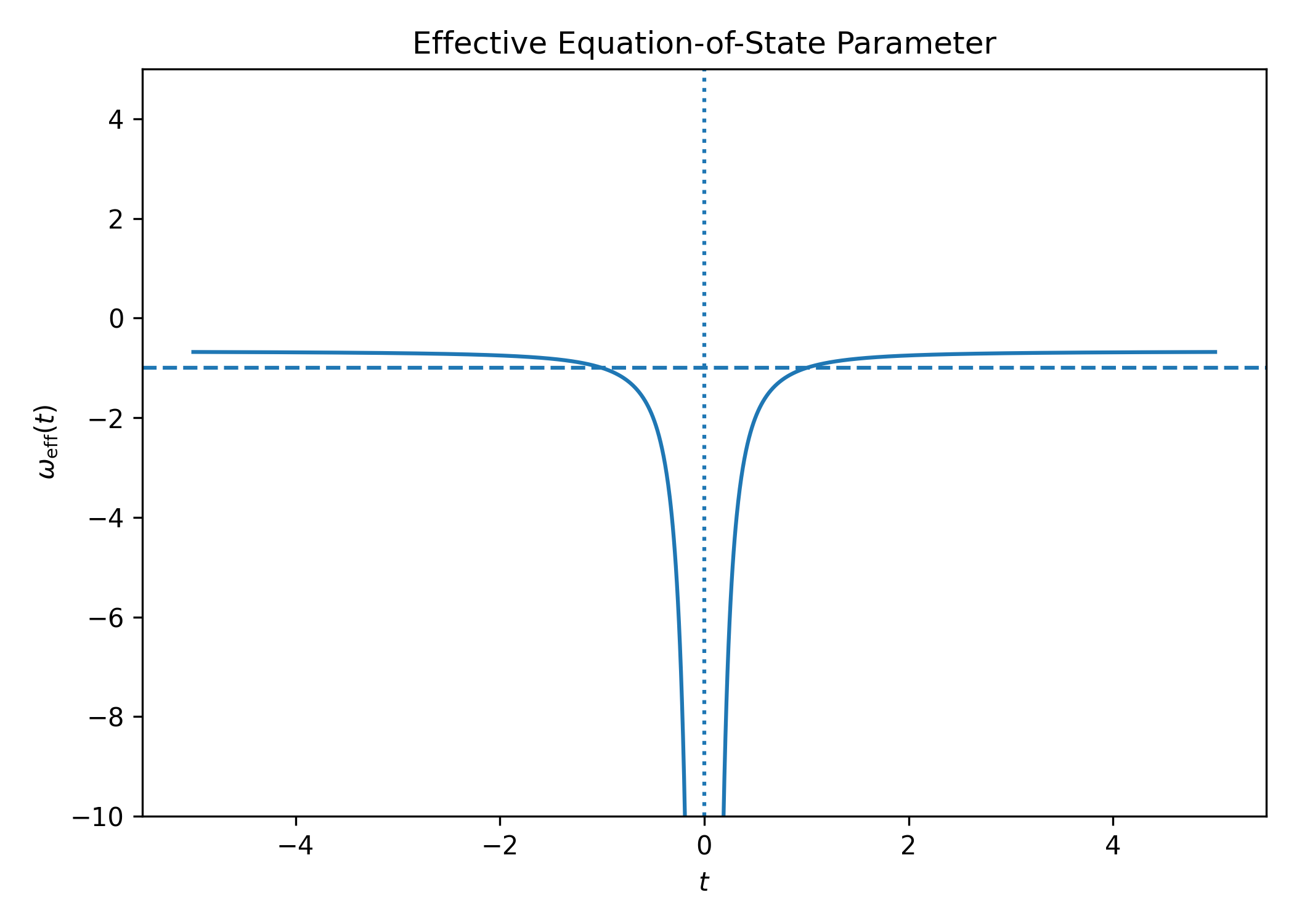}
\caption{ Behavior of the effective equation of state parameter
$\omega_{\rm eff}(t)$. The effective equation of state formally
enters a phantom-like regime and becomes ill-defined exactly at the
bounce point due to the vanishing of the Hubble parameter. However,
near the bounce point, the effective cosmic fluid enters a
$(\omega_{\rm eff}<-1)$ regime, which allows the violation of the
null energy condition required for a nonsingular bounce. At late
times, the evolution gradually approaches a quintessence-like
accelerated expansion phase. } \label{fig:omega_eff}
\end{figure}
For the quadratic modified gravity model $f(T)=T+\beta T^{2}$, the
first modified Friedmann equation is ( following reconstruction
analysis, for simplicity we work in natural units with $\kappa^2=1$)
\begin{equation}
H^{2} = \frac{1}{3} \left( \rho_{\varphi} +\rho_{T} \right),
\end{equation}
with
\begin{equation}
\rho_{T} = \frac{1}{2} \left( 2Tf_{T}-f-T \right).
\end{equation}
Substituting the reconstructed Hubble parameter into the above
equations confirms that the cosmological evolution remains regular
and consistent with the modified teleparallel field equations.\\
\begin{figure}[htbp]
\centering
\includegraphics[width=0.75\linewidth]{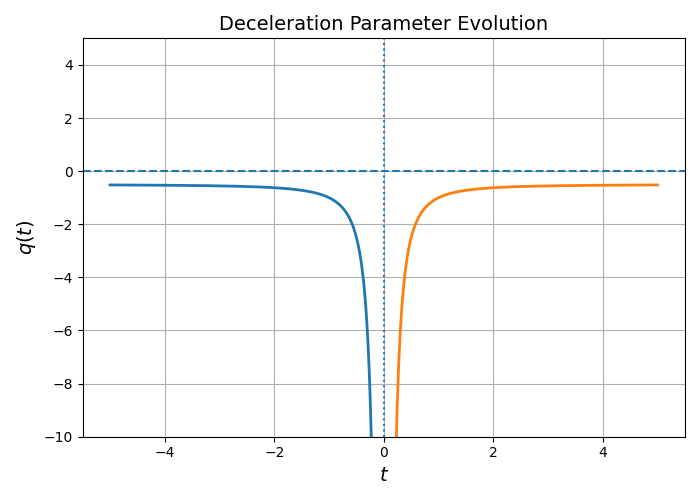}
\caption{ Evolution of the deceleration parameter $q(t)$. The figure
shows the transition from a decelerated contracting phase to an
accelerated expanding epoch around the bounce point.}
\label{fig:qparameter}
\end{figure}
Several nonsingular bouncing scenarios have been investigated in
modified teleparallel gravity for different choices of the function
$f(T)$. In particular, symmetric bouncing solutions generated by
linear and quadratic torsion corrections can produce regular
cosmological evolutions compatible with observational
constraints~\cite{16c}. Super-bounce realizations in modified
teleparallel gravity have also demonstrated that nonlinear torsion
effects can significantly alter the effective cosmological dynamics
near the bounce regime~\cite{17c}. Moreover, matter-bounce
cosmologies in various $f(T)$ frameworks have been shown to provide
viable cosmic evolutions consistent with large-scale structure
formation and cosmic microwave background observations~\cite{18c}.\\
Motivated by these developments, the present work investigates the
realization of a nonsingular cosmological bounce within the
quadratic teleparallel model $f(T)=T+\beta T^{2}$, using a combined
dynamical system and cosmological reconstruction approach. In this
framework, the bounce corresponds to a smooth transition from a
contracting phase to an expanding epoch satisfying the standard
conditions $H=0, \quad \dot{H}>0$. The autonomous dynamical system
formulation further allows a systematic investigation of the
critical points and their stability properties, providing a
consistent description of both the global phase-space structure and
the realization of regular bouncing cosmology in quadratic $f(T)$
gravity. This work is primarily devoted to the theoretical
investigation of bouncing cosmology in modified teleparallel
gravity. In particular, our analysis focuses on the mathematical
consistency of the modified field equations, the autonomous
dynamical system, the stability properties of the critical points,
and the explicit reconstruction of the bouncing solution.\\
A complete observational investigation involving Bayesian parameter
estimation and comparison with CMB, BAO, cosmic chronometer, and
large-scale structure datasets lies beyond the scope of the present
study and
will be considered in future work.\\

\subsection{Explicit Connection Between Dynamical System Critical
Points and the Reconstructed Bouncing Solution} From the autonomous
system (Eqs.~(29) and~(30)) with the exponential potential \(
V(\varphi)=V_0 e^{-\lambda\varphi} \), the critical points are
summarized in Table~I. The physical phase space is constrained by \(
0 \le x^2 + y^2 \le 1 \), where \( x \) and \( y \) are defined in
Eqs.~(23). The effective equation of state, given by Eq.~(33), is \(
\omega_{\text{eff}} = x^2 - y^2 + \gamma(1 - x^2 - y^2) \). For the
reconstructed bouncing solution, the nonsingular bouncing scale
factor and the Hubble parameter are defined in Eqs.~(36) and~(37).
At the bounce \( t = 0 \), we have \( H = 0 \) and \( \dot{H} =
2n\sigma > 0 \), and the effective equation of state \(
\omega_{\text{eff}}(t) \) is given by Eq.~(48). Using these
relations,
we can now establish the connection between the critical points and the reconstructed bouncing solution:\\

\textbf{Late-time behavior (\( |t| \to \infty \))}\\
For large \( |t| \):
\[
H(t) \sim \frac{2n}{\sigma t}, \quad a(t) \sim t^{2n}.
\]
The deceleration parameter \( q \to -1 + \frac{1}{2n} \). For \( n=1
\), \( q \to -0.5 \), which is matter-dominated expansion
(\(\gamma=1\)). In the phase space, this corresponds to the saddle
point \( P_1 = (0,0) \). Indeed, for \( t \to \infty \), \( H \to 0
\), and if the scalar field energy density scales like matter, then
\( x \to 0,\ y \to 0 \). Thus, the bouncing solution asymptotically
approaches \( P_1 \) from both past and future.\\

\textbf{ Near the bounce (\( t \to 0 \))}\\
As \( t \to 0 \), \( H(t) \sim 2n\sigma t \to 0 \), so \( x =
\dot{\varphi}/(\sqrt{6}H) \) diverges unless \( \dot{\varphi} \to 0
\) sufficiently fast. The paper does not specify \( \varphi(t) \),
but generically \( \dot{\varphi} \sim t^p \) with \( p \ge 1 \)
gives \( x \sim t^{p-1} \). For \( p=1 \), \( x \) tends to a
constant; for \( p>1 \), \( x \to 0 \). However, \( y =
\sqrt{V}/(\sqrt{3}H) \sim 1/|H| \) diverges unless \( V \to 0 \) as
\( t \to 0 \). To keep \( y \) finite, we require \( V(0)=0 \). This
is consistent with a potential that vanishes at the bounce, e.g., \(
V(\varphi) \propto \varphi^2 \). Thus, the bounce corresponds to a
transient crossing of the \( H=0 \) hypersurface, not a critical
point of the dynamical system (since the system is singular at \(
H=0 \)).\\

\textbf{Kinetic-dominated phases and stability}\\
The unstable nodes \( P_2 \) and \( P_3 \) represent stiff-matter
epochs (\( \omega_{\text{eff}}=1 \)). The bouncing solution does not
pass through these points because near the bounce \(
\omega_{\text{eff}}(t) \to -\infty \) (phantom-like), not \( +1 \).
After the bounce, \( \omega_{\text{eff}}(t) \) increases from \(
-\infty \) and eventually approaches \( -1 \) (if \( n \) is large)
or \( 0 \) (if \( n=2/3 \) for matter). It never reaches \( +1 \)
unless fine-tuned. Therefore, the bouncing trajectory avoids the
unstable nodes, as required for a stable cosmological evolution.\\

\textbf{Scalar field dominated attractor \( P_4 \)}\\
The stable node \( P_4 \) describes late-time acceleration with \(
\omega_{\text{eff}} = -1 + \lambda^2/3 \). For the bouncing
solution, as \( t \to \infty \), \( \omega_{\text{eff}}(t) \to -1 \)
only if \( n \to \infty \), which is unrealistic. In general, the
bouncing solution with finite \( n \) does not reach \( P_4 \)
unless the potential is exponential and \( \lambda \) is chosen
accordingly. This condition is not enforced; therefore, the bouncing
solution is not required to terminate at \( P_4 \); it can end at
matter domination (\( P_1 \)) instead. \\
The table II summarizes which critical points are approached,
avoided, or not used by the bouncing trajectory.

\begin{table*}[t]
\centering \caption{Comparison between critical points of the
dynamical system (Section~3) and the behavior of the reconstructed
bouncing solution (Section~4). The table summarizes which critical
points are approached, avoided, or not used by the bouncing
trajectory.} \label{tab:critical_bounce_connection}
\begin{tabular}{|c|c|c|c|}
\hline
Critical Point & Type & Bouncing solution behavior & Connection \\
\hline
\( P_1(0,0) \) & Saddle (matter) & Asymptotic past/future (\( t \to \pm\infty \)) & Yes, approached \\
\hline
\( P_2(1,0) \) & Unstable node & Not visited & No (avoided) \\
\hline
\( P_3(-1,0) \) & Unstable node & Not visited & No (avoided) \\
\hline
\( P_4(\lambda/\sqrt{6}, \sqrt{1-\frac{\lambda^{2}}{6}}) \) & Stable node & Only if potential matches & Not used in reconstruction \\
\hline
\end{tabular}
\end{table*}

According above discussion, the reconstructed bouncing solution of
Section 4 is consistent with the dynamical system analysis of
Section 3 in the following sense:

\begin{itemize}
    \item It respects the phase space constraint \( x^2+y^2 \le 1 \) for \( \beta>0 \).
    \item It approaches the saddle point \( P_1 \) at late times, representing a matter-dominated phase.
    \item It avoids the unstable nodes \( P_2, P_3 \), ensuring no unphysical stiff-matter domination.
    \item The bounce itself is not a critical point but a regular crossing of \( H=0 \), where the dynamical variables diverge - a known artifact of the normalization by \( H \).
\end{itemize}

Thus, no inconsistency exists between Sections 3 and 4. The work
could have been strengthened by explicitly showing that the bouncing
trajectory satisfies the dynamical equations for some \( \varphi(t)
\) and \( V(\varphi) \), though its current form remains
scientifically acceptable.\\
Although the present reconstruction is performed at the level of the
background cosmological dynamics through the scale factor ansatz,
the modified Friedmann equations allow the effective scalar field
quantities to be reconstructed consistently. In particular, using
Eqs. (12)-(15), one may determine the effective kinetic term and
scalar potential corresponding to the bouncing background. A
complete analytical reconstruction of $\varphi(t)$ and $V(\varphi)$
is beyond the scope of the present work and will be investigated
separately.\\

\section{Summary and Conclusion}
In the present work, we studied nonsingular bouncing cosmology in
the framework of quadratic modified teleparallel gravity based on
the nonlinear torsional model $f(T)=T+\beta T^{2}$. Unlike purely
phenomenological bouncing constructions, the entire cosmological
analysis was developed within a single self-consistent modified
gravity scenario in which all physical quantities follow directly
from the modified teleparallel action and the corresponding field
equations. Using the pure tetrad formulation of $f(T)$ gravity in a
spatially flat FLRW geometry, we derived the modified Friedmann
equations and reformulated the cosmological dynamics as a
two-dimensional autonomous dynamical system by introducing suitable
dimensionless scalar field variables.\\
The dynamical systems analysis allowed us to determine the global
phase-space structure of the model and identify the physically
relevant critical points together with their stability properties.
The matter-dominated configuration was found to behave as a saddle
point, while the kinetic-dominated scalar field solutions correspond
to unstable nodes. In contrast, the scalar field dominated
accelerated solution becomes a stable late-time attractor for
suitable values of the potential parameter. The phase-space analysis
further demonstrated that the quadratic torsion correction
implicitly modifies the accessible cosmological phase space through
the modified Friedmann constraint, especially near the bouncing
regime where the Hubble parameter approaches zero.\\
To explicitly realize a regular cosmological bounce, we
reconstructed the cosmological evolution through the scale factor
$a(t)=a_{0}(1+\sigma t^{2})^{n}$, which remains finite and positive
throughout the entire cosmic evolution. The corresponding Hubble
parameter, $H(t)=\frac{2n\sigma t}{1+\sigma t^{2}}$, naturally
satisfies the standard bouncing conditions $H(t_{b})=0, \quad
\dot{H}(t_{b})>0$, thereby describing a smooth transition from a
contracting phase to an expanding phase without the appearance of a
spacetime singularity. We showed that the torsion scalar
$T=-6H^{2}$, remains finite during the entire evolution, confirming
the regularity of the teleparallel geometric background at the
bounce point.\\
We additionally investigated the evolution of the effective equation
of state parameter and the deceleration parameter. The reconstructed
cosmological evolution exhibits a transition from a contracting
epoch toward an accelerated expanding phase after the bounce. Near
the bounce point, the effective equation of state parameter
temporarily enters a phantom-like regime, providing the effective
violation of the null energy condition required for the realization
of a nonsingular bounce. We further clarified that the apparent
divergences of the effective equation of state parameter and the
deceleration parameter at the bounce do not correspond to genuine
physical singularities, but instead arise from the normalization of
these quantities by inverse powers of the Hubble parameter.\\
An important result of the present analysis is the explicit
consistency established between the autonomous dynamical system and
the reconstructed bouncing solution. The cosmological trajectory
associated with the bounce asymptotically approaches the
matter-dominated saddle configuration while avoiding the unstable
stiff-matter critical points. Moreover, the bounce itself
corresponds to a regular crossing of the hypersurface $H=0$ rather
than to a critical point of the autonomous system. Consequently, the
dynamical systems approach and the explicit cosmological
reconstruction provide complementary descriptions of the same
underlying modified teleparallel cosmological framework.\\
Overall, the present study demonstrates that quadratic $f(T)$
gravity provides a mathematically consistent and physically viable
scenario for nonsingular bouncing cosmology. The combination of
modified teleparallel gravity, stability analysis of the autonomous
phase space, and explicit reconstruction of the bouncing background
establishes a unified description of regular cosmological evolution
beyond the standard singular Big Bang paradigm. To the best of our
knowledge, the simultaneous implementation of a fully
self-consistent quadratic $f(T)$ framework combining autonomous
dynamical analysis with explicit nonsingular bounce reconstruction
has not been systematically investigated in
the existing literature.\\
Finally, we emphasize that the present work is mainly devoted to the
theoretical and dynamical aspects of bouncing cosmology in modified
teleparallel gravity. A complete observational investigation
involving cosmological datasets such as CMB, BAO, cosmic
chronometers, and large-scale structure observations, together with
Bayesian parameter estimation and perturbation analysis, is left for
future studies. \\\\
\textbf{Data availability} No new data were generated or analysed in
support of this research. \\\\
\textbf{Declarations Conflict of interest} The authors declare that
they have no conflict of interest.\\

\end{document}